\documentclass[]{article}
\usepackage[letterpaper, margin=1in]{geometry}
\usepackage {graphicx}
\usepackage{amssymb}
\usepackage{amsmath}
\usepackage{amsbsy}
\usepackage{setspace}
\usepackage{natbib}
\usepackage{color}
\usepackage{bm}
\usepackage{textcomp}
\usepackage{hyperref}

\newcommand{\mbf}[1]{\mbox{\boldmath${#1}$}}

\newcommand{\Z}{{\mbox{\boldmath $Z$}}}

\newcommand{\be}{\mbf{\beta}}

\newcommand{\bvarphi}{{\mbox{\boldmath $\bvarphi$}}}

\begin{document}

\title{Cautionary note on ``Semiparametric modeling of grouped current duration data with preferential reporting''}
\author{}
\author{Alexander C. McLain,$^{*1}$ Rajeshwari Sundaram,$^2$ Marie Thoma,$^3$ and Germaine M. Buck Louis$^4$}
\footnotetext[1]{\emph{Corresponding Author:} Department of Epidemiology and Biostatistics, Arnold School of Public Health, University of South Carolina, 450 Discovery I, 915 Greene Street, Columbia, SC 29208 USA.  Email: mclaina@mailbox.sc.edu}
\footnotetext[2]{Division of Intramural Population Health Research,\emph{Eunice Kennedy Shriver} National Institute of Child Health \& Human Development, National Institutes of Health.}\footnotetext[3]{Department of Family Science,School of Public Health, University of Maryland.}\footnotetext[4]{College of Health and Human Services, George Mason University.}

\date{}
\maketitle

\maketitle

\doublespacing

\section{Introduction}
This report is designed to clarify a few points about the article ``Semiparametric modeling of grouped current duration data with preferential reporting'' by McLain, Sundaram, Thoma and Louis in \emph{Statistics in Medicine} \citep[][hereafter MSTL]{McLetal14} regarding using the methods under right censoring.  In simulation studies, it has been found that bias can occur when right censoring is present.  Current duration data normally does not have censored values, but censoring can be induced at a value, say $\tau$, after which the data values are thought to be unreliable.  As noted in MSTL, some right censored data require an assumption on the parametric form of the data beyond $\tau$.  While this assumption was given in MSTL, the implications of the assumption were not sufficiently explored.  Here we present simulations and evaluate the methods of MSTL under type I censoring, give some settings under which the method works well even in presence of censoring, state when the model is correctly specified and discuss the reasons of the bias.

\section{Tail Assumptions Under Right Censoring}

The bias observed under censoring is a result of model misspecification under censoring.  To see this, we note the following form of the current duration probability mass function for the semi-parametric model
\begin{eqnarray}\label{eq.discrete.CD}
g(y|\Z) = \frac{\exp\left\{-\exp(\be^\top \Z)\sum_{j=0}^y \alpha_j\right\} }{\sum_{y=0}^\infty \exp\left\{-\exp(\be^\top \Z)\sum_{j=0}^y \alpha_j\right\}},
\end{eqnarray}
where $\alpha_j \geq 0$ for all $j$ with $\alpha_0\equiv 0$.  When there is no censoring, the denominator in (\ref{eq.discrete.CD}) is calculated by setting $\alpha_y = \infty$ for $y>Y_{(m)}$ where $Y_{(m)}$ is the maximum observed current duration.  The infinite sum in (\ref{eq.discrete.CD}) then stops at $Y_{(m)}$.  However, such an approach cannot be taken under right  censoring. As noted in the Estimation section of MSTL, page 3966,
\begin{quote}
Let  $\tilde Y_{(1)}, \tilde Y_{(2)}, \ldots,  \tilde Y_{(m)} \leq \tau$ denote the ordered and distinctly observed uncensored current durations, and $\bar{G}(y|\Z) = 1-\sum_{j=0}^y g(j|\Z)$.   When censoring is present, we cannot set $\alpha_y = \infty$ for $y>\tilde Y_{(m)}$ because the likelihood for those censored at $\tau$ would be $\bar{G}(\tau|\Z)=0$.  To allow for $\bar{G}(\tau|\Z)>0$, we introduce an additional parameter $\alpha_{\tau}$ and set $\alpha_y = \alpha_{\tau}$ for all $y >\tilde Y_{(m)}$.  
\end{quote}
That is, under type I censoring the model assumes that $\alpha_j$ are equal for all $j\geq \tau$. 

The tail assumption is needed because a semiparametric model cannot estimate the mean under type I censoring without making a parametric assumption on the distribution beyond the value of $\tau$.  Recall that the relationship between $Y$ and $T$ is $\bar{F}_T(y) = g(y)\mu_T$, thus $\mu_T = E(T)$ is required to specify the model.  Under type I censoring at $\tau$ we can only estimate $E(T|T\leq \tau)$ with a semiparametric model.  This is similar to the fact that $\mu_T$ cannot be estimated from a Kaplan-Meier curve if the maximum value is censored.  To estimate $\mu_T$ the above tail assumption is used, which implies that the discrete hazard probability of $T$ takes the parametric form
$$ \lambda(y|\Z) = P(T = y|T \geq y, \Z) = 1 - \exp\{-\alpha_{\tau^+} \exp(\be^\top \Z)\} \mbox{ for all  }y\geq \tau.$$
Notice that this implies that the discrete hazard probabilities are constant in $y$, thus $T$ follows a geometric distribution in the tail, i.e., $\lambda(y|\Z)$ is constant in $y$ for $y\geq \tau$.  When this assumption is misspecified biases can occur.  For example, if $\lambda(y|\Z)$ is non-constant in $y$ for $y\geq \tau$, the denominator in (\ref{eq.discrete.CD}) is misspecified since it is a function of $\lambda(y|\Z)$ for $y\geq \tau$.  The misspecification in the denominator cannot be absorbed in any way, and results in model misspecification. This same phenomena happens with the piecewise constant model of MSTL, where $\alpha_y$ is constant beyond the largest knot.

If the values of $T$ were observed the tail behavior of the $\alpha_j$'s would not impact the estimation since they would not enter the likelihood.  However, since we observe the $Y$ values with probability mass function given in (\ref{eq.discrete.CD}), the tail values of $\alpha_j$ impact the estimation.  This explains why this problem is unique to current duration analysis.

Another issue with censoring is how to truncate the upper limit of the infinite sum in the denominator in (\ref{eq.discrete.CD}), which we denote by $Y^+$.  In theory this value should be set at a point where negligible probability mass occurs thereafter.  For cases when there is no known upper boundary to the distribution, we have observed in simulation studies that when $Y^+$ is too large it causes instability in the estimates, especially for the piecewise constant model, and having $Y^+$ too small results in biased estimates.  Whether a value is ``too small'' or ``too large'' will depend on the distribution of the data. A strategy we found effective in simulation studies was to set $Y^+$ to twice the largest value before the administrative censoring was implemented.  MSLT set $Y^+=1000$, which we found could be too large based on some of the new simulation settings tested.

\section{Simulation Studies}

To test the properties of the models in MSLT, numerous simulation studies were performed.  The current duration for the $i$th subject was simulated by generating the unobserved total durations as $T_{ij} {\sim} F$ for $j = 1,2,\ldots,K$, where $K = \min(k; \sum_{j=1}^k T_{ij}>M)$ and $M$ is a fixed large integer then setting $Y_{ij} = T_{iK} - M$.  This setting replicates a renewal process in equilibrium with renewal distribution \citep[see][for details]{Fel66}.

All of the simulation scenarios used data that was discretely distributed with a simple binary covariate $X$ with 0.5 success probability.  The underlying distribution of the survival times is $P(T=t|T\geq t) = 1 - \exp\{-\alpha_t\exp(\beta_1 X)\}$ where $\beta_1=0.5$.  The value of $\alpha_t$ was set to (a) $\alpha_t=\theta$, (b) $\alpha_t=\theta \alpha_0 t_k^{\alpha_0-1}$ for $t \in (t_{k-1},t_k]$ or (c) $\alpha_t=\theta\{t^{\alpha_0} - (t-1)^{\alpha_0}\}$.  Here, (a) corresponds to a geometric setting, (b) corresponds to a piecewise geometric distribution, and the survival function for (c) is equal to $\bar F(t|X)=\exp\{-\alpha_0^t\exp(\beta_1 X)\}$ with we refer to as the discrete Weibull setting (note that (c) is equivalent to (a) when $\alpha_0=1$).  For (a) we set $\theta=1/5$, for (b) and (c) $\alpha_0=4/5$ and $\theta$ was varied to alter the proportion of censored values.  For (b) $\theta=3/16$ or $\theta=3/8$, while for (c) $\theta=1/4$ or $\theta=1/8$.  The lower $\theta$ values induce more censoring.  For (b) we set $\{t_1,\ldots,t_7\}=\{1, 2, 4, 5, 7, 10,18\}$ and $t_0=0$, which match the knots used for the piecewise constant model.  For each  setting, type I censoring at $\tau = \{3,6,12,24,36\}$ along with no censoring was applied.  All simulations used $n=1000$ subjects.

The above distributions were fitted with the semiparametric and piecewise constant models from MSLT where the piecewise constant model had knots at $\{1, 2, 4, 5, 7, 10,18\}$, equal to those used for simulating the data.  For the geometric setting in (a) the tail assumption is correctly specified regardless of the value of $\tau$.  The tail assumption is also correctly specified in (b) when $\tau \geq 18$ since $\alpha_j=\alpha_{18}$ for all $j \geq 18$.  The misspecified scenarios include (b) when $\tau < 18$, and setting (c).  Programs to simulate and fit all models are available from the first authors website (see the `Programs' Section below).

\begin{table}[!t]
\caption{Summary of $1,000$ simulated samples with $n=1000$ for the piecewise constant and semi-parametric models under various discrete distributional assumptions with fixed type I censoring at $\tau$. Displayed is the true coefficient (\textsc{true}), the average estimated coefficient (\textsc{mean}), the empirical bias (\textsc{bias}), the empirical standard deviation (\textsc{sd}), the empirical coverage probability (\textsc{ecp}) and the censoring proportion (\textsc{prop cen}).}
\begin{center}
\begin{tabular}{lccccccccccc} \hline \hline
 & &\multicolumn{4}{c}{Piecewise Constant} && \multicolumn{4}{c}{Semi-parametric} & \\ & &\multicolumn{9}{c}{Geometric}  \\			
$\tau$	&	\textsc{true}	&	\textsc{mean}	&	\textsc{bias}	&	\textsc{sd}	&	\textsc{ecp}	&	&	\textsc{mean}	&	\textsc{bias}	&	\textsc{sd}	&	\textsc{ecp}	&	\textsc{prop cen}	\\
3	&	0.5	&	0.496	&	-0.004	&	0.082	&	0.959	&	&	0.498	&	-0.002	&	0.087	&	0.945	&	0.365	\\
6	&	0.5	&	0.505	&	0.005	&	0.077	&	0.951	&	&	0.498	&	-0.003	&	0.081	&	0.949	&	0.178	\\
12	&	0.5	&	0.503	&	0.003	&	0.075	&	0.946	&	&	0.499	&	-0.001	&	0.075	&	0.949	&	0.039	\\
24	&	0.5	&	0.504	&	0.004	&	0.074	&	0.947	&	&	0.509	&	0.009	&	0.073	&	0.954	&	0.003	\\
36	&	0.5	&	0.507	&	0.007	&	0.073	&	0.947	&	&	0.509	&	0.009	&	0.073	&	0.950	&	0.001	\\
None	&	0.5	&	0.507	&	0.007	&	0.073	&	0.958	&	&	0.509	&	0.009	&	0.073	&	0.950	&	0.000	\\
 & &\multicolumn{9}{c}{Piecewise Geometric}  \\
$\tau$	&	\textsc{true}	&	\textsc{mean}	&	\textsc{bias}	&	\textsc{sd}	&	\textsc{ecp}	&	&	\textsc{mean}	&	\textsc{bias}	&	\textsc{sd}	&	\textsc{ecp}	&	\textsc{prop cen}	\\
3	&	0.5	&	0.591	&	0.091	&	0.077	&	0.809	&	&	0.580	&	0.080	&	0.079	&	0.864	&	0.453	\\
6	&	0.5	&	0.567	&	0.067	&	0.071	&	0.869	&	&	0.561	&	0.061	&	0.074	&	0.906	&	0.281	\\
12	&	0.5	&	0.538	&	0.038	&	0.062	&	0.926	&	&	0.527	&	0.027	&	0.065	&	0.944	&	0.134	\\
24	&	0.5	&	0.508	&	0.008	&	0.058	&	0.957	&	&	0.509	&	0.009	&	0.059	&	0.956	&	0.055	\\
36	&	0.5	&	0.509	&	0.009	&	0.057	&	0.960	&	&	0.514	&	0.014	&	0.059	&	0.950	&	0.025	\\
None	&	0.5	&	0.509	&	0.009	&	0.057	&	0.959	&	&	0.516	&	0.016	&	0.059	&	0.951	&	0.000	\\
 & \multicolumn{11}{c}{Piecewise Geometric with high censoring}  \\			
$\tau$	&	\textsc{true}	&	\textsc{mean}	&	\textsc{bias}	&	\textsc{sd}	&	\textsc{ecp}	&	&	\textsc{mean}	&	\textsc{bias}	&	\textsc{sd}	&	\textsc{ecp}	&	\textsc{prop cen}	\\
3	&	0.5	&	0.705	&	0.205	&	0.117	&	0.573	&	&	0.689	&	0.189	&	0.120	&	0.672	&	0.752	\\
6	&	0.5	&	0.654	&	0.154	&	0.098	&	0.659	&	&	0.636	&	0.136	&	0.102	&	0.764	&	0.631	\\
12	&	0.5	&	0.587	&	0.087	&	0.079	&	0.808	&	&	0.555	&	0.055	&	0.081	&	0.930	&	0.472	\\
24	&	0.5	&	0.506	&	0.006	&	0.060	&	0.947	&	&	0.484	&	-0.016	&	0.067	&	0.928	&	0.311	\\
36	&	0.5	&	0.506	&	0.006	&	0.058	&	0.941	&	&	0.488	&	-0.012	&	0.064	&	0.924	&	0.209	\\
None	&	0.5	&	0.506	&	0.006	&	0.057	&	0.938	&	&	0.510	&	0.010	&	0.056	&	0.950	&	0.000	\\
 & &\multicolumn{9}{c}{Discrete Weibull}  \\			
$\tau$	&	\textsc{true}	&	\textsc{mean}	&	\textsc{bias}	&	\textsc{sd}	&	\textsc{ecp}	&	&	\textsc{mean}	&	\textsc{bias}	&	\textsc{sd}	&	\textsc{ecp}	&	\textsc{prop cen}	\\
3	&	0.5	&	0.548	&	0.048	&	0.088	&	0.930	&	&	0.540	&	0.040	&	0.092	&	0.946	&	0.571	\\
6	&	0.5	&	0.523	&	0.023	&	0.080	&	0.946	&	&	0.520	&	0.020	&	0.084	&	0.938	&	0.395	\\
12	&	0.5	&	0.511	&	0.011	&	0.070	&	0.948	&	&	0.504	&	0.004	&	0.074	&	0.942	&	0.200	\\
24	&	0.5	&	0.503	&	0.003	&	0.066	&	0.937	&	&	0.500	&	0.000	&	0.068	&	0.934	&	0.058	\\
36	&	0.5	&	0.503	&	0.003	&	0.065	&	0.943	&	&	0.508	&	0.008	&	0.068	&	0.937	&	0.018	\\
None	&	0.5	&	0.502	&	0.002	&	0.065	&	0.942	&	&	0.510	&	0.010	&	0.067	&	0.938	&	0.000	\\
 & \multicolumn{11}{c}{Discrete Weibull with high censoring}  \\						
$\tau$	&	\textsc{true}	&	\textsc{mean}	&	\textsc{bias}	&	\textsc{sd}	&	\textsc{ecp}	&	&	\textsc{mean}	&	\textsc{bias}	&	\textsc{sd}	&	\textsc{ecp}	&	\textsc{prop cen}	\\
3	&	0.5	&	0.570	&	0.070	&	0.123	&	0.932	&	&	0.555	&	0.055	&	0.125	&	0.943	&	0.783	\\
6	&	0.5	&	0.551	&	0.051	&	0.101	&	0.930	&	&	0.538	&	0.038	&	0.103	&	0.949	&	0.661	\\
12	&	0.5	&	0.536	&	0.036	&	0.084	&	0.936	&	&	0.517	&	0.017	&	0.091	&	0.943	&	0.481	\\
24	&	0.5	&	0.523	&	0.023	&	0.070	&	0.952	&	&	0.497	&	-0.003	&	0.077	&	0.941	&	0.267	\\
36	&	0.5	&	0.519	&	0.019	&	0.066	&	0.955	&	&	0.493	&	-0.007	&	0.072	&	0.949	&	0.154	\\
None	&	0.5	&	0.518	&	0.018	&	0.064	&	0.954	&	&	0.515	&	0.015	&	0.066	&	0.949	&	0.000	\\ \hline \hline
\end{tabular}
\end{center}
\label{SimComp1}
\end{table}

In Table \ref{SimComp1} we present bias, standard deviation and empirical coverage probabilities for various distributional assumptions corresponding to the distributions discussed above, which were varied by the fixed censoring value and the $\theta$ and $\alpha_0$ parameters.  As expected, the effect of the varying censoring value on the geometric setting is relatively small.  There does appear to be a decrease in the overall parameter estimate as the censoring value decreases, but overall the estimates are relatively unbiased.  For the piecewise geometric setting the parameters are relatively unbiased for $\tau \geq 24$. This is as hypothesized since when $\tau \geq 24$ the tail assumption is correctly specified.  When $\tau < 24$ the tail assumption is misspecified and we see increasing bias as $\tau$ gets closer to zero.  Further, when the proportion censored increases the results remain consistent.  This suggests that the value of $\tau$, not the overall censoring proportion, is what is driving the bias.  Thus, when the tail assumption is correctly specified the results appear to be relatively unbiased regardless of the proportion censored. 

The Weibull setting shows noticeable bias in the estimates when the censoring percentage is larger than 10\%.  It should be noted that the piecewise constant model is misspecified under the Weibull, so some bias is expected.  This misspecification appears to have a larger impact on the bias for the `high censoring' distribution.  For the semi-parametric setting the results have small bias when the censoring proportion is less than 30\%.

\section{Discussion}

The purpose of this paper was to investigate the properties of the MSTL model when all data are censored at a fixed value (i.e., type I censoring at $\tau$).  The impact of censoring is that a parametric assumption on the tail behavior of the data must be assumed.  Specifically, under censoring the model assumes that the hazard probability is constant for all $y\geq \tau$ where $\tau$ is the censoring value.  The simulation studies show that when the tail behavior is correctly specified both models have relatively unbiased results regardless of the amount of censoring.  This can be seen in the relatively unbiased results for the geometric setting for both models (another setting with higher censoring showed similar results), and the results for both piecewise geometric settings when $\tau >18$.  Recall that the last knot of the piecewise scenario was  $18$ so the true $\alpha_j$ values are constant beyond this value. Thus, when $\tau >18$ the distribution is geometric beyond the censoring value. The discrete Weibull setting is misspecified for all values of $\tau$.  Further, the piecewise constant model is misspecified when there is no censoring.  Our simulation results show that under misspecification the degree of bias depends on the amount of censoring.

The analysis included in MSTL censored all values at $\tau=36$.  The simulation studies suggest that $\tau=36$ will not have large impact on the results, however, this could be sensitive to the true distribution.  The analysis was repeated without censoring and the results were largely unchanged.  The previous analysis with the piecewise model found significant associations for both age ($\beta = -0.035$ with 95\% CI $[-0.054, -0.015]$) and parity ($\beta = -0.492$ with 95\% CI $[0.257, 0.728]$).  This analysis also found significant associations for both age ($\beta = -0.037$ with 95\% CI $[-0.053, -0.021]$) and parity ($\beta = 0.470$ with 95\% CI $[0.272, 0.667]$).  For the semi-parametric model the effect of age changed from $\beta = -0.036 \ [-0.074, 0.003]$ in the old analysis to $\beta = -0.040 \ [-0.057, -0.022]$ with no censoring.  The effect of parity showed attenuation with $\beta = 0.747 \  [0.206, 1.289]$ in the old analysis and $\beta = 0.472 \ [0.272, 0.672]$ with the new analysis.

The ``geometric in the tail'' assumption allows calculation of the necessary quantities needed to implement maximum likelihood estimation under censoring.  Specifically, it assures that $\bar{G}(\tau|\Z)>0$ for all $\Z$ which is required for likelihood calculation.  When the ``geometric in the tail'' assumption is misspecified it will lead to biased results of varying degrees (as explored Section 3).  When the tail assumption is misspecified, one option is to impose different tail behavior.  Some examples include (i) $\alpha_t = \alpha_{\tau^+}^{\gamma(t-\tau-1)}$,  (ii) $\alpha_t =\theta\{t^{\alpha_0} - (t-1)^{\alpha_0}\}$, or (iii) $\alpha_t = (t-\tau-1)^\gamma$ for $t\geq \tau$.  It is important to keep in mind that sparse data are available to determine the tail behavior.  We implemented different tail assumptions in simulations studies and found unstable results when two parameters were included in the calculation of the tail behavior of $\alpha_j$.  So if (i) or (ii) were used one of the parameters should be fixed.

In summary, the simulations in the paper show that censoring should be employed with caution when using the MSTL method.  Further, if censoring is required multiple values of $\tau$ should be used to test the sensitivity of the results.  Unlike the situation found in standard survival analysis, the model assumptions extend beyond the censoring value. The main reason for censoring in current duration data is due to concerns  of measurement errors associated with large responses. Censoring is an attractive option when measurement error is likely, but we recommend that it be used cautiously in keeping with the specified parametric assumptions. One solution in this case is to use the piecewise model, which as shown in MSTL can correct for random digit preference in the outcome.

\section*{Software}

A zip file containing all the programs to implement the MSTL model can be found at through the following link \href{https://sites.google.com/site/alexmclain/research}{\color{blue} https://sites.google.com/site/alexmclain/research\color{black}}.  See the link under the reference for MSTL ``Zip file with R code to run the programs.''  This file contains all of the programs to run the semiparametric and piecewise models, along with a nonparametric method.  It also contains sample data, along with two programs that will generate current duration data for the discrete Weibull and piecewise constant distributions used in Section 3. The geometric distribution can be generated as a special case of the discrete Weibull distribution when $\alpha=1$.

\section*{Acknowledgements}

We would like to thank Professor Niels Keiding and his group for alerting us to this issue.

\bibliographystyle{asa}
\bibliography{Reference}

\end{document}